\documentclass[reprint,superscriptaddress,nofootinbib,aps,prl,onecolumn,notitlepage]{revtex4-1}
\usepackage[a4paper, margin=2.50cm]{geometry}
\usepackage{bbm}
\usepackage{graphicx}
\usepackage{fancyhdr, color}
\usepackage[english]{babel}

\usepackage[T1]{fontenc}
\usepackage[latin9]{inputenc}
\usepackage{color}
\usepackage{lipsum}
\usepackage{amsthm}
\usepackage{amsmath}
\usepackage{amssymb,mathtools}
\usepackage{graphicx}
\usepackage{array,tabulary} 
\usepackage[unicode=true,pdfusetitle, bookmarks=true,bookmarksnumbered=false,bookmarksopen=false,  breaklinks=false,pdfborder={0 0 0},backref=false,colorlinks=false] {hyperref}
\hypersetup{ colorlinks,linkcolor=myurlcolor,citecolor=myurlcolor,urlcolor=myurlcolor}
\usepackage{txfonts}
\usepackage{colortbl}\definecolor{myurlcolor}{rgb}{0,0,0.7}

\makeatletter
\@ifundefined{textcolor}{}
{%
 \definecolor{BLACK}{gray}{0}
 \definecolor{WHITE}{gray}{1}
 \definecolor{RED}{rgb}{1,0,0}
 \definecolor{GREEN}{rgb}{0,1,0}
 \definecolor{BLUE}{rgb}{0,0,1}
 \definecolor{CYAN}{cmyk}{1,0,0,0}
 \definecolor{MAGENTA}{cmyk}{0,1,0,0}
 \definecolor{YELLOW}{cmyk}{0,0,1,0}
 }

\newcolumntype{K}[1]{>{\centering\arraybackslash}m{#1}}

\theoremstyle{plain}
\newtheorem{thm}{\protect\theoremname}
\theoremstyle{plain}

\ifx\proof\undefined
\newenvironment{proof}[1][\protect\proofname]{\par
\normalfont\topsep6\p@\@plus6\p@\relax
\trivlist
\itemindent\parindent
\item[\hskip\labelsep
\scshape
#1]\ignorespaces
}{%
\endtrivlist\@endpefalse
}
\providecommand{\proofname}{Proof}
\fi
\theoremstyle{plain}
\newtheorem{lem}[thm]{\protect\lemmaname}

\newtheorem{defi}[thm]{\protect\definitionname}

\providecommand{\lemmaname}{Lemma}
\providecommand{\theoremname}{Theorem}
\providecommand{\definitionname}{Definition}
\providecommand{\propositionname}{Proposition}

\newcommand{\tr}{{\operatorname{Tr\,}}}
\newcommand{\id}{{\mathbb{I}}}
\newcommand{\e}{{\operatorname{e}}}
\newcommand{\dd}{{\operatorname{d}}}

\newcommand{\nc}{\newcommand}
\nc{\rnc}{\renewcommand}

\nc{\cT}{{\cal T}}
\nc{\1}{{\openone}}
\nc{\ox}{\otimes}
\nc{\cH}{{\cal H}}
\nc{\cM}{{\cal M}}
\DeclareMathOperator*{\argmin}{arg\,min}

\def\proj#1{| #1 \rangle\!\langle #1 |}
\nc{\ketbra}[2]{|#1\rangle\!\langle#2|}
\nc{\avg}[1]{\langle#1\rangle}

\newcommand{\norm}[1]{\left\| #1 \right\|}

\newcommand{\chrto}{{\color[rgb]{0.4,0.2,0.9} \sc Qbook:Ch.25}} 

\begin{document}

\fancyhead[C]{\sc \color[rgb]{0.4,0.2,0.9}{Quantum Thermodynamics book}}
\fancyhead[R]{}

\title{Thermodynamics from information}

\author{ Manabendra Nath Bera}
\email{mnbera@gmail.com}
\affiliation{ICFO -- Institut de Ciencies Fotoniques, The Barcelona Institute of Science and Technology, ES-08860 Castelldefels, Spain}
\affiliation{Max-Planck-Institut f\"ur Quantenoptik, D-85748 Garching, Germany}

\author{Andreas Winter}
\affiliation{Departament de F\'isica: Grup d'Informaci\'o Qu\`antica, Universitat Aut\`onoma de  Barcelona, ES-08193 Bellaterra (Barcelona), Spain}
\affiliation{ICREA, Pg.~Lluis Companys 23, ES-08010 Barcelona, Spain}

\author{Maciej Lewenstein}
\affiliation{ICFO -- Institut de Ciencies Fotoniques, The Barcelona Institute of Science and Technology, ES-08860 Castelldefels, Spain}
\affiliation{ICREA, Pg.~Lluis Companys 23, ES-08010 Barcelona, Spain}

\date{\today}

\begin{abstract}

Thermodynamics and information have intricate inter-relations. The justification of the fact that {\it information is physical}, is done by inter-linking information and thermodynamics -- through Landauer's principle. This modern approach towards information recently has improved our  understanding of thermodynamics, both in classical and quantum domains. Here we show thermodynamics as a consequence of information conservation. Our approach can be applied to most general situations, where systems and thermal-baths could be quantum, of arbitrary sizes and even could posses inter-system correlations. The approach does not rely on an {\it a priori} predetermined temperature associated to a thermal bath, which is not meaningful for finite-size cases. 
Hence, the thermal-baths and systems are not different, rather both are treated on an equal footing. This results in a ``temperature''-independent formulation of thermodynamics. We exploit the fact that, for a fix amount of \emph{coarse-grained} information, measured by the von Neumann entropy, any system can be transformed to a state that possesses minimal energy, without changing its entropy. This state is known as a {\it completely passive} state, which assumes Boltzmann--Gibb's canonical form with an {\it intrinsic temperature}. This leads us to introduce the notions of bound and free energy, which we further use to quantify heat and work respectively. 
With this guiding principle of information conservation, we develop universal notions of equilibrium, heat and work, Landauer's principle and also universal fundamental laws of thermodynamics. We show that the maximum efficiency of a quantum engine, equipped with a finite baths, is in general lower than that of an ideal Carnot's engine. We also introduce a resource theoretic framework for {\it intrinsic-temperature} based thermodynamics, within which we address the problem of work extraction and state transformations. 


\end{abstract}

\maketitle

\thispagestyle{fancy}

\section{Introduction}
Thermodynamics, being one of the most basic foundations of modern science, not only plays an important role in modern technologies, but also provides  basic understanding of the vast range of natural phenomena. Initially, it was phenomenologically developed to address the issues related to heat engines, i.e., the question on how, and to what extent, heat could be converted into work. But, with the developments of statistical mechanics, relativity and quantum mechanics, thermodynamics has attained quite a formal and mathematically rigorous form \cite{Gemmer09} along with its fundamental laws. It plays important roles in understanding relativistic phenomena in astrophysics and cosmology, in microscopic systems with quantum effects, or in complex systems in biology and chemistry.

The inter-relation between information and thermodynamics \cite{Parrondo15} has been studied in the context of Maxwell's demon \cite{Maxwell08, Leff90, Leff02, Maruyama09}, Szilard's engine \cite{Szilard29}, and Landauer's principle \cite{Landauer61, Bennett82, Plenio01, Rio11, Reeb14}. 
The classical and quantum information theoretic approaches help us to explain thermodynamics for small systems \cite{Shannon48, Nielsen00, Cover05}.  Recently, information theory has played an important role to explore thermodynamics with inter-system and system-bath correlations \cite{Alicki13, Marti15, Bera16}, equilibration processes \cite{Short11, Goold16, Rio16, Gogolin16}, and foundational aspects of statistical mechanics \cite{Popescu06} in quantum mechanics. Inspired by information theory, a resource theoretic framework for thermodynamics \cite{Brandao13} has been developed, which can reproduce standard thermodynamics in the asymptotic (or thermodynamic) limit. For small systems or in the finite copy limit (also known as one-shot limit), it reveals that the laws of thermodynamics require modifications to dictate the transformations on the quantum level \cite{Dahlsten11, Aberg13, Horodecki13, Skrzypczyk14, Brandao15, Cwiklinski15, Lostaglio15, Egloff15, Lostaglio15a}.

In the following, we make an axiomatic construction of thermodynamics by elaborating on the inter-relations between information and thermodynamics, and identify the ``information conservation'', measured by the von Neumann entropy, as the crucial underlying property for that. This has been formulated recently in \cite{BeraPRX}. 
We introduce the notion of \emph{bound energy}, which is the amount of energy locked in a system that cannot be accessed (extracted) given a set of allowed operations. We recover standard thermodynamics as a spacial case, by assuming ({\it i}) global \emph{entropy preserving (EP) operations} as the set of allowed operations and ({\it ii}) infinitely large thermal baths initially uncorrelated from the system.

All fundamental physical theories share a common property, that is information conservation. It implies that the underlying dynamics deterministically and bijectively map the set of possible configurations of a system between any two instants of time. A non-deterministic feature, in a classical world,  appears due to ignorance of some degree of freedoms, leading  to an apparent information loss. In contrast, this loss of information could be intrinsic in quantum mechanics, and occurs in measurement processes and in presence of non-local correlations \cite{BeraPhilo16}.  The set of entropy preserving operations is larger than the set of unitary operations in the sense that
they conserve entropy (coarse-grained information) but not the probabilities (fine-grained information). This is why we refer \emph{coarse-grained} information conservation. 
Both coarse-grained and fine-grained information conservation become equivalent (see Sec.~\ref{sec:EPoperations}) in the asymptotic limit. Note that unitaries are the only linear operations that are entropy preserving for all states \cite{Hulpke2006}. Except few specific example, To what extent a general coarse-grained information conserving operations can be implemented in the single-copy limit is yet to be understood. 

Note that an alternative approach to tackle thermodynamics for finite quantum systems relies fluctuation theorems.
The second law is obtained there as a consequence of reversible transformations on initially thermal states or states with a well defined temperature \cite{Jarzynski2000, Esposito09, Sagawa2012}. In contrast, the aim of our work is instead to  generalize thermodynamics, that is valid for arbitrary environments, irrespective of being thermal, or considerably larger than the system. This is illustrated  in the table below. Our formalism, which treats systems and environments on equal footing, results in a ``temperature''-independent formulation of thermodynamics. 

\begin{center}
{\small
\begin{tabular}{K{2cm}K{2.8cm}K{2.8cm}}
& {\bf Unitaries (fine-grained IC$^*$)}  & {\bf EP operations (coarse-grained IC)} \\
\hline
{\bf Large thermal bath}  & Resource theory of Thermodynamics & Standard Thermodynamics \\
\hline
{\bf Arbitrary environment} & ? & \hspace{.3cm }Our formalism \\
\hline
\end{tabular}}
\end{center}
\vspace{-.25cm}\hspace{.3cm}
{\footnotesize *IC: Information Conservation}

\vspace{.3cm}
This ``temperature''-independent thermodynamics is essential in contexts where the environment and the system are comparable in size, or the environment simply not being thermal. In the real experimental situations, environments are not necessarily thermal, but can even possess quantum properties, like coherence or correlations.

The entropy preserving operations allow us to represent all the states and thermodynamic processes
in a simple energy-entropy diagram. It shows that all the states with equal energies and entropies thermodynamically equivalent.
We give a diagrammatic representation for heat, work and other thermodynamic quantities, and exploit a geometric approach to understand their transformation under thermodynamics processes. This could enable us to reproduce several results in the literature, for example resource theory of work and heat in \cite{Sparaciari16}, and also to extend thermodynamics involving multiple conserved quantities \cite{BeraPRX} in terms of generalized Gibbs ensembles.

\section{Entropy preserving operations, entropic equivalence class and intrinsic temperature}
\label{sec:EPoperations}
We consider \emph{entropy preserving (EP) operations} that arbitrarily change an initial state $\rho$ without changing its entropy
\begin{equation*}
\rho \to \sigma \hspace{.3cm}: \hspace{.3cm} S(\rho)=S(\sigma)\, ,
\end{equation*}
where $S(\rho)\coloneqq - \tr(\rho \log \rho)$ is the von Neumann entropy. Sometime we denote these operations as \emph{ iso-informatic} operations as well. Note, in general, these operations are not linear operations, i.e., an operation that acts on $\rho$ and produces a state with the same entropy, not necessarily preserves entropy when acting on other states. As was shown in Ref. \cite{Hulpke2006}, a quantum channel $\Lambda(\cdot)$ that preserves entropy as well as linearity, i.~e.\ $\Lambda(p \rho_1+ (1-p)\rho_2)=p\Lambda(\rho_1)+ (1-p)\Lambda(\rho_2)$, has to be unitary.

Given any two states $\rho$ and $\sigma$ with $S(\rho)=S(\sigma)$, and an ancillary system $\eta$ of $O(\sqrt{n\log n})$ qubits, there exists a global unitary $U$ such that \cite{Sparaciari16}
\begin{equation}\label{eq:entropy-pres-micro}
\lim_{n\to \infty}
\|\tr_{\textrm{anc}}\left(U \rho^{\otimes n}\otimes\eta U^{\dagger}\right)-\sigma^{\otimes n}\|_1=0  \, ,
\end{equation}
where the partial trace is performed on the ancillary system and $\|\cdot\|_1$ is the one-norm. The reverse statement is also true, i.e. if two states respect Eq.~\eqref{eq:entropy-pres-micro}, then they also have equal entropies. 

For thermodynamics, it is important to restrict entropy preserving operations that are also be  energy preserving. These operations can also be implemented. In Theorem 1 of Ref.~\cite{Sparaciari16}, it is shown that if two states $\rho$ and $\sigma$ have equal entropies and energies, i.e. ($S(\rho)=S(\sigma)$ and $E(\rho)=E(\sigma)$), then there exists energy preserving $U$ and an additional ancillary system $A$ in some state $\eta$ with $O(\sqrt{n\log n})$ of qubits and Hamiltonian $\norm{H_A}\leqslant O(n^{2/3})$, for which \eqref{eq:entropy-pres-micro} is fulfilled.
Note, in the large $n$ limit, the amount of energy and entropy of the ancillary system per copy vanishes.

Let us introduce different equivalence classes of states, depending on their entropy, by which we establish a hierarchy of states depending on their information content. 
\begin{defi}[Entropic equivalence class]
For any quantum system of dimension $d$, two states $\rho$ and $\sigma$  are equivalent 
and belong to the same entropic equivalence class if and only if
both have the same Von Neumann entropy,
\begin{equation*} 
\rho \sim \sigma \hspace{1cm} \textrm{iff} \hspace{1cm} S(\rho)=S(\sigma)\, .
\end{equation*}
\end{defi}
Assuming some fixed Hamiltonian $H$, the representative element of every class is the state that \emph{minimizes} the energy within it, i.e.,
\begin{equation*}
\gamma(\rho)\coloneqq\argmin_{\sigma \ : \ S(\sigma)=S(\rho)} E(\sigma) ,
\end{equation*}
where $E(\sigma)\coloneqq \tr(H \sigma)$ is the energy of the state $\sigma$.

Complementary to the maximum-entropy principle \cite{Jaynes57a, Jaynes57b}, that identifies the thermal state as the state that maximizes the entropy for a given energy, one can show that, the thermal state also minimizes the energy for a given entropy. We refer to this latter property as \emph{min-energy principle} \cite{Pusz78, Lenard78, Alicki13}, which identifies thermal states as the representative elements of every class as 
\begin{equation}\label{eq:thermal-state}
\gamma(\rho) = \frac{\e^{-\beta(\rho) H}}{\tr \left(\e^{-\beta(\rho) H}\right)}\,.
\end{equation}
The $\beta (\rho)$ is inverse temperature and labels the equivalence class, to which the state $\rho$ belongs. This $\beta (\rho)$ is denoted as the ``intrinsic'' inverse temperature associated to $\rho$. The state $\gamma (\rho)$ is, also, referred to as the completely passive (CP) state \cite{Alicki13} with the minimum internal energy, but with the same information content. These CP states, with the form $\gamma(H_S, \beta_S)$, has  several interesting properties \cite{Pusz78, Lenard78}: (P1) It minimizes the energy, for a given entropy. (P2) With the decrease (increase) in $\beta_S$, both energy and entropy monotonically increase (decrease), and vice versa. (P3) For non-interacting Hamiltonians, $H_T=\sum_{X=1}^N \mathbb{I}^{\otimes X-1} \otimes H_X \otimes \mathbb{I}^{\otimes N-X}$, the joint complete passive state is tensor product of individual ones, i.e, $\gamma(H_T, \beta_T)=\otimes_{X=1}^N \gamma(H_X, \beta_T)$, with identical $\beta_T$ \cite{Pusz78, Lenard78}.

\section{Bound and free energies: energy-entropy diagram}
Here we identify two relevant forms of internal energy: the free and the bound energy. Indeed, the notions depend on the set of allowed operations. While the latter is defined as the amount of internal energy that is accessible in the form of work. For entropy preserving operations, in which the entropic classes and CP states arise, it is quantified as in the following.
\begin{defi}[Bound energy \cite{BeraPRX}]
\label{defi:bound-energy}
The bound energy for a state $\rho$ with the system Hamiltonian $H$ is
\begin{equation}\label{eq:bound-energy-def}
 B(\rho)\coloneqq  \min_{\sigma \ :  \ S(\sigma)=S(\rho)} E(\sigma)=E(\gamma(\rho)),
\end{equation}
where $\gamma(\rho)$ is the state with minimum energy (CP), within the equivalence class of $\rho$.
\end{defi}
As guaranteed by the min-energy principle, $B(\rho)$ is the amount of energy that cannot be extracted further by performing any entropy preserving operations. Amount of bound energy is strongly related to the information content in the state. Only by allowing an outflow of information from the system, one could have access to this energy (in the form of work).

On other hand, the free energy is the accessible part of the internal energy:
\begin{defi}[Free energy]
\label{defi:free-energy}
The free energy stored in a system $\rho$, with system Hamiltonian $H_S$, is given by
\begin{equation}\label{eq:free-energy-def}
F(\rho) \coloneqq E(\rho)-B(\rho),
\end{equation}
where $B(\rho)$ is the bound energy in $\rho$.
\end{defi}

Note, the free energy does not rely on a predefined temperature, in contrast to the standard out of equilibrium Helmholtz free energy $F_T(\rho)\coloneqq E(\rho)-T\, S(\rho)$, where  $T$ is the temperature of a thermal bath. The situation in which the free energy in Eq.~\eqref{eq:free-energy-def} becomes equal to the accessible Helmholtz free energy is considered in Lemma~\ref{lem:work-extraction}.  Nevertheless, with intrinsic temperature $T(\rho)\coloneqq \beta(\rho)^{-1}$ that labels the equivalence class that contains $\rho$, our definition of free energy identifies with the out of equilibrium free energy as
\begin{equation*}
F(\rho)=
F_{T(\rho)}(\rho)-F_{T(\rho)}\left(\gamma(\rho)\right).
\end{equation*}
Let us mention that, in the rest of the manuscript, we use $F_{\beta}(\rho)\coloneqq E(\rho)-\beta^{-1}\, S(\rho)$ to denote the standard out of equilibrium free energy, wherever we find it more convenient. 

Beyond single systems, the notions of bound and free energy can be extended to multi-particle (multipartite) systems, where these quantities exhibit several interesting properties. They even can capture the presence of inter-party correlations. For bipartite system, the list of properties is given below.
\begin{lem} [\cite{BeraPRX}]
For a bipartite system, with non-interacting Hamiltonian $H_A\otimes \id + \id \otimes H_B$, in an arbitrary state $\rho_{AB}$ and a product state $\rho_A\otimes \rho_B$ with marginals $\rho_{A/B}\coloneqq \tr_{B/A}(\rho_{AB})$, the bound and the free energy satisfy the following:
\begin{enumerate}
\item[(P4)] Bound energy and correlations:
\begin{equation}\label{eq:bound-energy-corr}
 B(\rho_{AB})\le B (\rho_A\otimes \rho_B)\, .
\end{equation}
\item[(P5)] Bound energy of composite systems:
\begin{equation}\label{eq:bound-energy-composite}
B (\rho_A\otimes \rho_B)\le B (\rho_A)+B (\rho_B)\, .
\end{equation}
\item[(P6)] Free energy and correlations:
\begin{equation}\label{eq:free-energy-corr}
F(\rho_A\otimes \rho_B)\le F(\rho_{AB})\, .
\end{equation}
\item[(P7)] Free energy of composite systems:
\begin{equation}\label{eq:free-energy-composite}
F(\rho_A)+F(\rho_B)\le F(\rho_A\otimes \rho_B)\, .
\end{equation}
\end{enumerate}
Eqs.~\eqref{eq:bound-energy-corr} and \eqref{eq:free-energy-corr} are saturated if and only if $A$ and $B$ are uncorrelated $\rho_{AB}=\rho_A\otimes\rho_B$, and Eqs.~\eqref{eq:bound-energy-composite} and \eqref{eq:free-energy-composite} are saturated if and only if $\beta (\rho_A)=\beta(\rho_B)$.
\end{lem}

\begin{figure}
\includegraphics[width=\columnwidth]{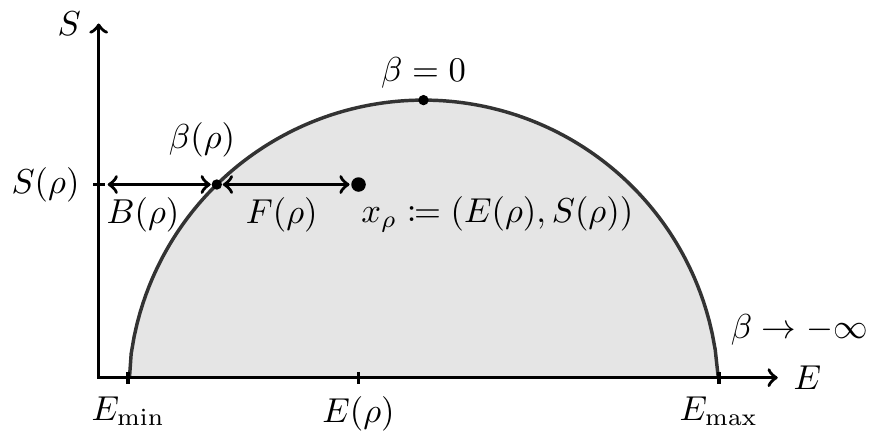}
\caption{
Energy-entropy diagram. Any quantum state $\rho$ is represented in the diagram
as a point with coordinates $x_\rho\coloneqq(E(\rho), S(\rho))$. Note, since we consider average energy and coarse-grained entropy as the coordinates, a single point in this diagram represents not one but many quantum states.   
The free energy $F(\rho)$, of the state $\rho$, can be seen from the diagram as the horizontal distance from the thermal boundary. This is, in fact, the part of internal energy which is accessible without altering system's entropy. The bound energy $B(\rho)$ is the horizontal distance  between the thermal boundary and the energy reference. Contrary to the free energy, it cannot be extracted by means of entropy preserving operations. The slope of the tangent line on a point in the thermal boundary represents the intrinsic temperature, $\beta(\rho)$, of the state $\rho$. 
\label{fig:energy-entropy-diagram}}
\end{figure}

These properties give an additional operational meaning to the free energy $F(\rho)$. For a system in a state $\rho$ and infinitely large bath at inverse temperature $\beta$, the extractable work under global entropy preserving operations is given by
\begin{equation}\label{eq:work-extracted-infinite-bath}
W= F_\beta (\rho) - F_\beta (\gamma(\beta))\, ,
\end{equation}
where the standard free energy is $F_\beta (\rho)=E(\rho)-\beta^{-1}S(\rho)$ and $\gamma(\beta)$ is the thermal state with 
the inverse temperature $\beta$, which is the resultant state once the work has been extracted.

\begin{lem}[Free energy vs. $\beta$-free energy \cite{BeraPRX}]
\label{lem:free-energy-beta}
The free energy $F(\rho)$ is the one that corresponds to work extracted by attaching a bath at the worst possible temperature,
\begin{equation*}
F(\rho)=\min_\beta \left(F_\beta (\rho) - F_\beta (\gamma(\beta))\right)\, .
\end{equation*}
The minimization is achieved when inverse temperature becomes identical to the inverse intrinsic temperature $\beta(\rho)$.
\end{lem}

\emph{Energy-entropy diagrams} are very useful in thermodynamics, where every point in the digram is represented by energy and entropy.  
As shown in Fig.~\ref{fig:energy-entropy-diagram}, a state $\rho$ of a system with Hamiltonian $H$ is represented by a point with coordinates $x_\rho\coloneqq(E(\rho), S(\rho))$. All physical states are bounded in a region made of the horizontal axis (i.e., $S=0$, corresponding to the pure states) and the convex curve $(E(\beta),S(\beta))$ correspond to the thermal states of both positive and negative temperatures which is denoted as the thermal boundary. The slope of the tangent line associated to a point on the thermal boundary is given by the inverse temperature, since $\frac{\dd S(\beta)}{\dd E(\beta)}=\beta$.

%

In general a point in the energy-entropy diagram could correspond to multiple states, as different quantum states can have identical entropy and energy. This conversely shows that the energy-entropy diagram establishes a link between the microscopic and the macroscopic thermodynamics, i.e. in the asymptotic limit all the thermodynamic quantities only rely on the energy and entropy per particle \cite{Sparaciari16}. In fact, all the states with equal entropy and energy are thermodynamically equivalent in the sense that they can be inter-converted into each other in the limit of many copies $n\to\infty$, using energy conserving unitary operations with an ancilla of sub-linear size $O(\sqrt{n \log n})$ and a Hamiltonian upper bounded by sub-linear bound $O(n^{2/3})$ \cite{Sparaciari16}.

\section{Temperature independent formulation of thermodynamics }

\subsection{Equilibrium and zeroth law\label{sec:zeroth}}
In thermodynamics, the zeroth law establishes the absolute scaling of temperature and the notion of thermal equilibrium. It states that if systems $A$ and $B$ are in mutual thermal equilibrium, $B$ and $C$ are also in mutual thermal equilibrium, then  $A$ will also be in thermal equilibrium with $C$. All these systems, in mutual thermal equilibrium, are classified to thermodynamically equivalent class where each state in the class is assigned with a unique parameter called temperature and there is no spontaneous energy exchange in between them. 
When a non-thermal state is brought in contact with a large thermal bath, the system may exchange energy and entropy to acquire a thermally equilibrium state with the temperature of the bath. This is also known as equilibration process, during which the system could exchange both energy and entropy with bath such that it minimizes its Helmholtz free energy. 

However, such an equilibration process would be very different when the system cannot have access to a considerably large thermal bath or in absence of a thermal bath. A formal definition of equilibration, in that case, based on information preservation and intrinsic temperature can be given as in the following.
\begin{defi}[Equilibrium and zeroth law \cite{BeraPRX}]
\label{def:equilibrium}
Given a collection of systems $A_1, \ldots, A_n$ with non-interacting Hamiltonians $H_1, \ldots H_n$ in a 
joint state, $\rho_{A_1 \ldots A_n}$, we call them to be mutually at equilibrium if and only if 
\begin{equation*}
F(\rho_{A_1 \ldots A_n})=0\, ,
\end{equation*}
i.e. they ``jointly'' minimize the free energy as defined in \eqref{eq:free-energy-def}.
\end{defi}

Two states $\rho_A$ and $\rho_B$, with corresponding Hamiltonians $H_A$ and $H_B$, achieve mutual equilibrium when they jointly attain an iso-informatic (i.e. iso-entropic) state with minimum possible energy. The resultant equilibrium state is indeed a completely passive (CP) state $\gamma(H_{AB}, \beta_{AB})$ with the joint (non-interacting) Hamiltonian $H_{AB}=H_A\otimes \mathbb{I} + \mathbb{I} \otimes H_B$, 
it is 
\begin{align}
 \gamma(H_{AB},\beta_{AB})=\gamma(H_A, \beta_{AB}) \otimes \gamma(H_B, \beta_{AB}),
\end{align}
where the local systems are also in completely passive states with the same $\beta_{AB}$. This follows from the property (P3). Note, we denote $\gamma(H_X, \beta_Y)=e^{-\beta_Y H_X}/\tr(e^{-\beta_Y H_X})$. 
As a corollary, it can be seen that if two CP states $\gamma(H_A, \beta_A)$ and $\gamma(H_B, \beta_B)$ are with $\beta_A \neq \beta_B$, then they can, still, jointly reduce bound-energy without altering the total information content and acquire their mutual equilibrium state $\gamma(H_{AB}, \beta_{AB})$. From the property (P5), this implies a unique $\beta_{AB}$, i.e., 
\begin{align}
\label{eq:eqlbAB}
E(\gamma(H_A, \beta_A))+E(\gamma(H_B, \beta_B)) > E(\gamma(H_{AB}, \beta_{AB})). 
\end{align}
Moreover for $\beta_A \geq \beta_B$, the equilibrium temperature $\beta_{AB}$ is bounded. This is expressed in Lemma \ref{lm:betaAB} below and has been proven in \cite{BeraPRX}.
\begin{lem}[\cite{BeraPRX}]
\label{lm:betaAB}
Iso-informatic equilibration process between $\gamma(H_A, \beta_A)$ and $\gamma(H_B, \beta_B)$, with $\beta_A \geq \beta_B$,
results in a mutually equilibrium joint state $\gamma(H_{AB}, \beta_{AB})$, where $\beta_{AB}$ satisfies
\begin{align}
 \beta_A \geq \beta_{AB} \geq \beta_B, \nonumber
\end{align}
irrespective of non-interacting systems' Hamiltonians.
\end{lem}

Now with the notion of equilibration and equilibrium state as the global CP state, we could recast the \emph{zeroth law}, in terms of intrinsic temperature. A global CP state assures that the individual states are not only CP states with vanishing inter-system correlations but also they share identical intrinsic temperature, i.e., $\beta$. In reverse, individual systems are in mutual equilibrium as long as they are locally in a CP state and share the same intrinsic temperature.

The traditional notion of thermal equilibrium, as well as, the zeroth law can also be recovered using the argument above. In traditional sense, the thermal baths are reasonably large, in comparison to the systems under consideration, with a predefined temperature. For a bath Hamiltonian $H_B$, a bath can be expressed as $\gamma_B=e^{-\beta_B H_B}/\tr(e^{-\beta_B H_B})$, where$\beta_B$ is the inverse temperature and $|\gamma_B|\rightarrow \infty$. Note it is also a CP state. When a finite system in a state $\rho_S$ (with $|\rho_S|\ll |\gamma_B|$) with Hamiltonian $H_S$ is brought in contact with a thermal bath, the global state, after reaching mutual thermal equilibrium, will be a CP state, i.e., $\gamma_B \otimes \rho_S \xrightarrow{\Lambda^{ep}} \gamma_B^\prime \otimes \gamma_S$, with a global inverse temperature $\beta_e$. It can be easily seen that $\gamma_B^\prime \rightarrow \gamma_B$, $\beta_e \rightarrow \beta_B$ and $\gamma_S \rightarrow e^{-\beta_B H_S}/\tr(e^{-\beta_B H_S})$, in the limit $|\gamma_B| \gg |\rho_S|$ and $|\gamma_B|\rightarrow \infty$. 

\subsection{Work, heat and the first law \label{sec:first}}
In thermodynamics, the conservation of energy is assured by the first law by taking into account the distribution of energy over work and heat, that are the two forms of energy transfer. 

Consider a thermodynamic process that involves a system $A$ and a bath $B$ and a transformation $\rho_{AB}\to \rho_{AB}'$ that respect conservation of the global entropy $S(\rho_{AB})=S(\rho_{AB}')$. Traditionally, a bath is by definition assumed to be initially thermal and fully uncorrelated from the system. Then the heat dissipation is usually quantified as the internal energy change in the bath, i.e. $\Delta Q=E(\rho_B')-E(\rho_B)$ where the reduced state of the bath is $\rho_B^{(\prime)}=\tr_A \rho_{AB}^{(\prime)}$. This definition, however, has been shown to have limitations.  A consistent definition has been discussed recently in \cite{Bera16} based on information theoretic approach. It has been suggested that heat has to be quantified as $\Delta Q=T_B \Delta S_B$, where $T_B$ being the temperature and $\Delta S_B= S(\rho_B')-S(\rho_B)$ is the von Neumann entropy change in the bath. This definition can further be generalized to the situation where the system and bath are correlated, where $\Delta S_B=-\Delta S(A|B)$ is also the conditional entropy change in system $A$, conditioned on the bath $B$, defined as $S(A|B)=S(\rho_{AB}) - S(\rho_B)$. Thus, in the presence of correlations, heat flow can be understood as the energy exchange due to information flow from the system to the bath.

However this definition will not be meaningful if the environment is athermal or not in the state of the Boltzmann-Gibbs form. For arbitrary environments, an alternative, and meaningful, quantification of heat exchange can still be given in terms of the change in bound energies.
\begin{defi}[Heat]
\label{def:heat}
For a system $A$ and environment $B$, the dissipated heat by the system $A$ in the process $\rho_{AB} \xrightarrow{\Lambda^{ep}} \rho_{AB}'$ is the change in bound energy of the environment $B$, i.e.
\begin{equation}\label{eq:heat-def}
\Delta Q \coloneqq B(\rho_B')-B(\rho_B).
\end{equation}
Here $B(\rho_B^{(\prime)})$ is the initial (final) bound energy of the bath $B$.
\end{defi}

Clearly heat is a process dependent quantity. There might be many different processes, that transform the same initial to the same final marginal state for $A$, but with different marginal states for $B$. Since the global process is entropy preserving, for same marginal states for $A$, the global process could lead to different entropy change for $B$ and also to a different amount of correlations between $A$ and $B$, which is measured by the mutual information $I(A:B)\coloneqq S_A+S_B-S_{AB}$. Note $\Delta S_A + \Delta S_B = \Delta I (A:B)$.

\begin{figure}
\includegraphics[width=0.8\columnwidth]{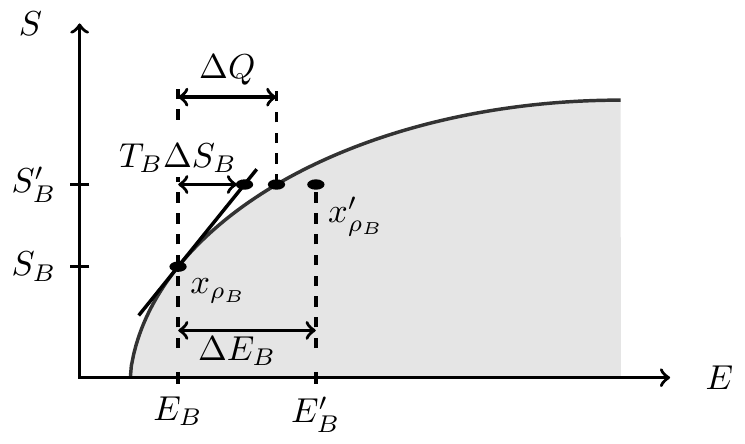}
\caption{
The different notions of heat can be understood using the energy-entropy diagram. They are (1) $\Delta Q$ as change of the bound energy of the bath, (2) $\Delta E_B$ as the change in internal energy, and (3) $T_B \Delta S_B$. In this example, the bath is initially thermal but this is not necessary.
\label{fig:heat-definitions}}
\end{figure}

%

As special cases, this quantification of heat can give rise to other existing definitions \cite{BeraPRX}.  In particular, for an initial thermal environment $\rho_B=\e^{-H_B/T}/\tr(\e^{-H_B/T})$ with Hamiltonian $H_B$ at temperature $T$, then the exchanged heat is upper and lower bounded by
\begin{equation}\label{eq:heat-defs-bounds}
T \Delta S_B\le \Delta Q \le \Delta E_B\, .
\end{equation}
These three quantities have been schematically represented in the energy-entropy diagram in Fig.~\ref{fig:heat-definitions}. The three definitions become equivalent in the limit of large thermal baths, as in that case, a small perturbation of the bath $\rho_B'=\rho_B+\delta \rho_B$ leads to 
\begin{equation}\label{eq:heat-dif-def}
 T \Delta S_B +O(\delta \rho_B^2)=\Delta Q=\Delta E_B-O(\delta\rho_B^2)\, ,
\end{equation}
and the second order contribution will vanish.
With this consistent definition of heat, let us turn to  define work.

\begin{defi}[Work]
\label{defi:work} 
For a system $A$ and its environment $B$, and an arbitrary entropy preserving transformation $\rho_{AB} \rightarrow \rho_{AB}'$, with fixed non-interacting Hamiltonians $H_A$ and $H_B$, the work performed on the system $A$ is quantified as,
\begin{equation*}
\Delta W_A\coloneqq W - \Delta F_B
\end{equation*}
where the work cost implement the global transformation is $W=\Delta E_A + \Delta E_B$ and $\Delta F_B= F(\rho_B')-F(\rho_B)$.
\end{defi}

Now equipped with the notions of heat and work, the first law takes the form of a mathematical identity.
\begin{lem}[First law]
\label{lm:1stLaw} 
For a system $A$ and its environment $B$ with fixed non-interacting Hamiltonians $H_A$ and $H_B$, and an arbitrary entropy preserving transformation $\rho_{AB} \rightarrow \rho_{AB}'$, the change in energy for system $A$ is distributed as
\begin{equation*}
 \Delta E_A = \Delta W_A - \Delta Q \, .
\end{equation*}
\end{lem}
This can be deduced directly from the definitions of work and heat. Recall, $-\Delta W_A$ is the amount of ``pure'' energy, i.e. work, and  $\Delta Q_A$ is the hear which is change in energy due to exchange of information with the system.

\subsection{Second law \label{sec:second}}
The second law of thermodynamics can be expressed in several forms. These include an upper bound on the extracted work, or the impossibility of complete conversion of heat into work etc. Below we show how all existing formulations are a consequence of the principle of information conservation.

\subsubsection{Work extraction}
In thermodynamics, one of the main concerns is to convert heat into work, which is the ``pure'' form of energy and can be used for any application. 
\begin{lem}[Work extraction]\label{lem:work-extraction}
The extractable work from an arbitrary composite system $\rho$, by using entropy preserving process $\rho \to \rho'$,
is upper-bounded by the free energy
\begin{equation*}
W \le F(\rho),
\end{equation*}
where $W=E(\rho)-E(\rho')$ and the equality is reached if and only if $\rho'=\gamma(\rho)$.

If the composite is a bipartite system with the form $\rho=\rho_A\otimes \gamma_B(T_B)$, where $\gamma_B(T_B)$ is thermal at temperature $T_B$, then
\begin{equation}\label{eq:recover-standard-free-energy}
W \le F_{T_B}(\rho_A) - F_{T_B}\left(\gamma_A(T_B)\right)
\end{equation}
where the standard out of equilibrium free energy is $F_T(\cdot)$.  The equality is achieved in the limit of asymptotically large bath (or with infinite heat capacity).
\end{lem}

The first part can be proven by seeing that $E(\rho')\geq B(\rho')$ and therefore $W=E(\rho)-E(\rho')\le E(\rho)-B(\rho')=F(\rho)$. Note, under iso-informative transformation, $B(\rho')=B(\rho)$. For bipartite composite case $\rho=\rho_A \otimes \gamma_B(T_B)$, we may write
\begin{equation*}
\begin{split}
F(\rho_A\otimes \gamma_B(T_B))&=E(\rho_A)+E(\gamma_B(T_B))\\
&-\left(B(\gamma_A(T_{AB}))+B(\gamma_B(T_{AB}))\right)
\end{split}
\end{equation*}
where $T_{AB}$ is the intrinsic temperature of the composite $\rho_A\otimes \gamma_B(T_B)$.
%
Recall that thermal states satisfy $E(\gamma)=B(\gamma)$. Now by reshuffling the terms, we recover the first law, i.e.
\begin{equation*}
F(\rho_A\otimes \gamma_B(T_B))=-\Delta E_A - \Delta Q_A
\end{equation*}
where we use $\Delta E_A=E(\gamma_A(T_{AB}))-E(\rho_A)$ and $\Delta Q_A=B(\gamma_B(T_{AB}))-B(\gamma_B(T_B))$. Also we have $\Delta Q\ge T_B \Delta S_B$ (see Eq.~\eqref{eq:heat-defs-bounds}). Since the whole process is entropy preserving and we restrict initial and final composite to be in product state forms, we have $\Delta S_A=-\Delta S_B$, and therefore,
\begin{equation}\label{eq:work-dif-free-energies}
F(\rho_A\otimes \gamma_B(T_B))\le -\Delta E_A - T_B \Delta S_B=-\left(\Delta E_A - T_B \Delta S_A\right)\,
\end{equation}
which proves \eqref{eq:recover-standard-free-energy}.
Again, in the infinitely large bath limit, the final intrinsic temperature $T_{AB}$ will become the bath temperature $T_B$ and also $\Delta Q = T_B \Delta S_B$. Then the equality in Eq.~\eqref{eq:work-dif-free-energies} is achieved.


Now we turn to the question if heat can be converted into work. Traditionally, 
answer to this question leads to various other formulations of the second law in thermodynamics. They are the Clausius statement, Kelvin-Planck statement and Carnot statement, to mention a few. Indeed, similar question can be put forward in the frame-work considered here, in terms of bound energy, where one may not have access to large thermal bath. 
The analogous forms of second laws that consider this question both qualitatively and quantitatively, and they are outlined below. 

\subsubsection{Clausius statement}
Second law puts fundamental bound on extractable work, as well as dictates the direction of state transformations. Lets us first elaborate on the bounds on extractable work and thereby put forward the analogous versions of second law in this framework. 

\begin{lem}[Clausius statement]
For two systems $A$ and $B$ in an arbitrary states and with intrinsic temperatures $T_A$ and $T_B$ respectively, any iso-informatic process 
satisfy the inequality
\begin{equation}\label{eq:generalized_Clausius}
(T_B -T_A) \Delta S_A \geqslant \Delta F_A + \Delta F_B + T_B \Delta I(A:B) - W\, ,
\end{equation}
where the change in the free energy of the body $A/B$ is $\Delta F_{A/B}$ and $\Delta I (A:B)$ is the change of mutual information before and after the process. The $W=\Delta E_A + \Delta E_B$ is the amount of external work, which is performed on the global state. Note for initially uncorrelated state ($I(A:B)=0$), systems $A$ and $B$ are thermal ($F_A=F_B=0$), and no external work being performed ($W=0$), we have 
\begin{equation}\label{eq:std_Clausius}
(T_B -T_A) \Delta S_A \geqslant 0\, 
\end{equation}
as a corollary. This implies that there exists no iso-informatic equilibration process whose {\bf sole} result is the transfer of heat from a cooler to a hotter body.
\end{lem}

The Clausius statement above can be proven as in the following. By definition, free and bound energies satisfy
\begin{equation}\label{eq:energy_balance}
W=\Delta F_A + \Delta F_B + \Delta Q_A + \Delta Q_B\, .
\end{equation}
Recall, the heat is defines as the change of bound energy of the environment. Also, increase in bound energy implies increase in entropy, i.e. $\textrm{sign} (\Delta B)=\textrm{sign} (\Delta S)$. Now from Eq.~\eqref{eq:heat-defs-bounds} and with $T_{A/B}$ as the initial intrinsic temperature of the systems $A/B$, we may write
\begin{equation*}
\Delta Q_A +\Delta Q_B \geqslant T_B \Delta S_B + T_A \Delta S_A\, .
\end{equation*}
As a result of total entropy conservation, the change in mutual information is then $\Delta I (A:B)=\Delta S_A+\Delta S_B$. Now putting this in Eq.~\eqref{eq:energy_balance} gives raise to Eq.~\eqref{eq:generalized_Clausius}. 


Note the standard Clausius statement (as in Eq.~\eqref{eq:std_Clausius}) can be ``apparently'' violated, due to three reasons and that can be recovered from the general one in Eq.~\eqref{eq:generalized_Clausius}. Either the process not being spontaneous, which means external work is performed $W>0$, or due to the presence initial free energy in the systems (i.e. $F_A\neq 0$ or/and $F_B\neq 0$) which is consumed during the process, or due to the presence of initial correlations, i.e. $I (A:B)\neq 0$), which could lead to $\Delta I (A:B) <0$ \cite{Bera16}. 

\subsubsection{Kelvin-Planck statement}
While Clausius statement says that spontaneously heat cannot flow from a cooler to a hotter body, the Kelvin-Plank formulation of second law restricts it further stating that the heat flowing from a hotter to a colder body, cannot be transformed into work completely. A generalized form of that we present below.
\begin{lem}[Kelvin-Planck statement]
For two systems $A$ and $B$ in arbitrary states, any iso-informatic process satisfies
\begin{equation}\label{eq:generalized_Kelvin-Planck}
\Delta Q_B+\Delta Q_A = -(\Delta F_A + \Delta F_B) + W ,
\end{equation}
where $\Delta F_{A/B}$ is the change in the free energy of the body $A/B$ before and after the process, $\Delta Q_{A/B}$ the heat exchanged by the body $A/B$. The $W=\Delta E_A + \Delta E_B$ is the amount of external work invested, on the global state, to carry out the process.

For the case where the reduced states are thermal and the process is a work extracting one $W<0$, then above equality reduces
\begin{equation}\label{eq:Kelvin-Planck}
\Delta Q_B+\Delta Q_A \leqslant  W < 0\, .
\end{equation}
Further, in absence of initial correlations, Eq.~\eqref{eq:Kelvin-Planck} implies that there exists no iso-informatic process whose {\bf sole} result is the absorption of bound energy (heat) from an equilibrium state and converting it into work completely.
\end{lem}
The Eq.~\eqref{eq:generalized_Kelvin-Planck} is followed from the energy balance \eqref{eq:energy_balance}. The Eq.~\eqref{eq:Kelvin-Planck} is deduced from \eqref{eq:generalized_Kelvin-Planck} by assuming reduced states that are initially in thermal states and therefore $\Delta F_{A/B}\geqslant 0$.
The final statement is derived by noting that any entropy preserving process applied on initially uncorrelated systems fulfills $\Delta S_A + \Delta S_B \geqslant 0$, which together with \eqref{eq:Kelvin-Planck} leads to $\textrm{sign}(\Delta Q_A) =-\textrm{sign}(\Delta Q_B)$.


\subsubsection{Carnot statement}
Another version of second law is based on highest possible work conversion efficiency in an ideal heat engine, also known as Carnot statement.  

Consider a heat engine consists of two heat baths $A$ and $B$, with different temperatures $T_A$ and $T_B$ respectively. A working body $S$ cyclically interacts with $A$ and $B$. There is no restriction on how the working body interacts with the baths $A$ and $B$, apart from the fact that     the working body is left in its initial state and also uncorrelated with the bath(s) at the end of every cycle. This is to guarantee that the working body only absorbs heat from a bath and releases to the other one, without changing itself the end of the cycle. 

Here, in contrast to standard situations, we go beyond the assumption that baths are infinitely large. Rather, we consider the possibility that a bath could be similar in size as the system, such that loss or gain of energy changes their (intrinsic) temperatures. Consider, two uncorrelated baths   $A$ and $B$ at equilibrium with temperatures $T_A$ and $T_B$ respectively, and $T_A<T_B$, i.~e.\ $\rho_{AB}=\gamma_A\otimes \gamma_B$. After one (or several complete) cycle(s) in the engine, the environments transformed to $\rho_{AB}\to \rho_{AB}'$. 

The efficiency of work extraction in a heat engine, defined as the fraction of energy that is taken from the hot bath and then transformed into work, is expressed as 
\begin{equation*}
\eta\coloneqq \frac{W}{-\Delta E_B}
\end{equation*}
where $W$ is the amount of work extracted from the heat absorbed $-\Delta E_B=E_B-E_B'>0$ from the hot environment. Below, we go on to give upper bounds on the efficiency for any heat engine consist of arbitrary baths.

\begin{lem}[Carnot statement]
For a heat engine working between two baths $\gamma_A \otimes \gamma_B$ each of which in a local equilibrium state with intrinsic temperatures $T_B > T_A$ and uncorrelated with each other, the bound on efficiency of work extraction is given by
\begin{equation}\label{eq:General_Carnot}
 \eta \leqslant 1 - \frac{\Delta B_A}{-\Delta B_B},
\end{equation}
where the change in bound energies of the systems $A$ and $B$ are $\Delta B_A$ and $\Delta B_B$ respectively.

In the special case where baths are considerably large and engine operates under global entropy preserving operations, the Carnot efficiency is recovered,
\begin{equation}\label{eq:Carnot}
\eta \leqslant 1 - \frac{T_A}{T_B}\, .
\end{equation}
\end{lem}

The statement above is respected for arbitrary baths, even for the ones with small sizes, and can be proven as follows. For a transformation $\rho_{AB}\to \rho_{AB}'$, the maximum extractable work is given by
\begin{equation*}
W=F(\rho_{AB})-F(\rho_{AB}')=(-\Delta E_B) - \Delta E_A>0\, .
\end{equation*}
Then the efficiency of conversion of heat into work becomes
\begin{equation*}
\eta = 1 - \frac{\Delta E_A}{-\Delta E_B}\, .
\end{equation*}
If $A$ being initially at equilibrium, then $\Delta F_A\geqslant 0$ and $\Delta E_A > \Delta B_A$, which is also true for $B$. As a consequence, 
$\eta\leqslant 1 - \frac{\Delta B_A}{-\Delta B_B}$, which is Eq.~\eqref{eq:General_Carnot}. In the large bath limit i.e. $\Delta B_A \lll B_A$, bound energy change becomes $\Delta B_A=T_A \Delta S_A$.
As a result, 
\begin{equation}\label{eq:Carnot-step}
\eta\leqslant 1 - \frac{T_A \Delta S_A}{-T_B \Delta S_B}\, .
\end{equation}
For globally entropy preserving operations, the joint entropy remains unchanged,  i.~e.\ $S_{AB}'=S_A+S_B$, and
 $\Delta S_A+\Delta S_B \geqslant 0$ or alternatively $\Delta S_A \geqslant -\Delta S_B$. Now this, together with \eqref{eq:Carnot-step}, implies \eqref{eq:Carnot}.

%

Note that efficiency, in \eqref{eq:General_Carnot}, is derived with the consideration that the initial bath states are thermal. While the final states may or may not thermal after first cycle. The equality is achieved if the final states, after first cycle, are also thermal where temperature  of the final states could be different form initial ones. If an engine cycle starts with non-thermal baths, the efficiency will be different from \eqref{eq:General_Carnot}, which has been discussed in details in \cite{BeraPRX}.  

\subsection{Third law}
In thermodynamics, the third law deals with the impossibility of attaining the absolute zero temperature. According to Nernst version, it states: ``it is impossible to reduce the entropy of a system to its absolute-zero value in a finite number of operations''. Very recently, the third law of thermodynamics has been proven in Ref.\ \cite{Masanes2017}, where it is shown to be a consequence of the unitarity character of the thermodynamic transformations. For example, let us consider the transformation that cools (erases) system $S$, initially in a state $\rho_S$, in the presence of a bath $B$
\begin{equation}\label{eq:erasure}
\rho_S\otimes \rho_B \ \ \to \ \ \proj{0}\otimes\rho_B', 
\end{equation}
where $\rho_B$ is a thermal state and  $\textrm{rank}(\rho_S)>1$. The dimension of the bath's Hilbert space, $d_B$, could be arbitrarily large but finite. Since bath $\rho_B$, by definition, is a full-rank state, the left hand side and the right hand side of Eq.~\eqref{eq:erasure} are of  
different ranks. Unitary operations preserve rank of of states. Therefore the transformation cannot be carried out using a unitary operation, irrespective of work supply, where the system attains an absolute zero entropy state.

In access to infinitely large baths and a sufficient work supply, the zero entropy state can only be produced. However, if one assumes a locality structure for the bath's Hamiltonian, such a cooling (unitary) process would take an infinitely large amount time. In case of finite dimensional bath and a finite amount of resources (e.~g.\ work, time), a quantitative bound on the achievable temperature can be given, as in \cite{Masanes2017}.

Note that the framework considered here replies on set of entropy preserving operations and they are more powerful than unitaries. Clearly, transformation \eqref{eq:erasure} is possible using entropy preserving operation, if
\begin{equation*}
S(\rho_S)\leqslant \log d_B - S(\rho_B),
\end{equation*}
and an access to work $W=F(\proj{0}\otimes\rho_B')-F(\rho_S\otimes \rho_B)$ to implement the operation. Since entropy preserving operations can be implemented by using a global unitaries acting on infinitely many copies (see Sec.~\ref{sec:EPoperations}), the absolute zero entropy state can be achieved by means of entropy preserving operations. This is in agreement with \cite{Masanes2017}, in the cases of infinitely large baths and unitary operations.

In conclusion, the third law of thermodynamics can be understood as a consequence of the microscopic reversibility (unitarity) of the transformation and is not necessarily respected by general entropy preserving operations.

\section{State Transformations: A temperature independent Resource Theory of Thermodynamics}
One of the important aspects this framework is that it can be exploited to provide a resource theory of thermodynamics, which is independent of temperature \cite{BeraPRX}. A similar formulation is also introduced in \cite{Sparaciari16}. Here we briefly introduce that. The main ingredients of every resource theory are the resourceless state space, which are of vanishing resource, and a set of of allowed state transformations. Here, CP states are resouceless and set of allowed operations, in general, are the ones that are energy non-increasing and entropy non-decreasing. 

Let us first restrict to the set of operations that are entropy preserving regardless the energy. Consider two states $\rho$ and $\sigma$, and  without loss of generality $S(\rho)\le S(\sigma)$. Then, there exists an $n$, for which
\begin{equation}\label{eq:equal-entropies}
S(\rho^{\otimes n})=S(\sigma^{\otimes m}).
\end{equation}
Therefore, there is an entropy preserving operation that transforms $n$-copies of $\rho$ to $m$-copies of $\sigma$, and vice versa. By such a trick, with access to arbitrary number of copies, the states $\rho$ and $\sigma$ with different entropies, can be brought to a same manifold of equal entropy. Note $\rho^{\otimes n}$ and $\sigma^{\otimes m}$ belong to the spaces of different dimension, and they can be made equal in dimensions by
\begin{equation}\label{eq:rho-to-sigma-times-0}
\Lambda \left( \rho^{\otimes n}\right) =\sigma^{m}\otimes \proj{0}^{n-m}.
\end{equation}
The $\rho^{\otimes n}$ and $\sigma^{\otimes m}\otimes \proj{0}^{n-m}$ live now in spaces of the same dimension. Eq.~\eqref{eq:rho-to-sigma-times-0} represents, in fact, a randomness compression process, in which the information in $n$-copies of $\rho$ is compressed to $m$-copies of $\sigma$, and $n-m$ systems are erased.
If we are restricted to only entropy preserving operations, the rate of 
transformation, from \eqref{eq:equal-entropies}, becomes
\begin{equation}\label{eq:rate-entropies}
r\coloneqq\frac{m}{n}=\frac{S(\rho)}{S(\sigma)}\, .
\end{equation}

In thermodynamics, however, energy also plays important roles and thus must be taken into account. Otherwise, it could be possible that the process considered in \eqref{eq:rho-to-sigma-times-0} is not favorable energetically if  $E(\rho^{\otimes n}) < E(\sigma^{\otimes m})$. Then, more copies of $\rho$ are required such that the transformation becomes energetically favorable. Further, it creates states (with non-zero entropy), $\phi$, such that
\begin{equation*}
\Lambda(\rho^{\otimes n})=\sigma^{\otimes m}\otimes \phi^{\otimes n-m}.
\end{equation*}
The operation satisfies the energy and entropy conservation constraints, therefore 
\begin{equation*}
\begin{split}
E(\rho^{\otimes n})&=E(\sigma^{\otimes m}\otimes \phi^{\otimes n-m}),\\
S(\rho^{\otimes n})&=S(\sigma^{\otimes m}\otimes \phi^{\otimes n-m})\, .
\end{split}
\end{equation*}
In other form, they represent a geometric equation
of the points $x_\psi=(E(\psi),S(\psi))$ with $\psi \in \{\rho,\sigma,\phi \}$
\begin{equation}\label{eq:convex-combination}
x_\rho = r \ x_\sigma + (1-r) \ x_\phi \, ,
\end{equation}
in the energy-entropy diagram and $r\coloneqq m/n$ is the state conversion rate. Here the extensivity of both entropy and energy in the number of copies, e.g.\ $E(\rho^{\otimes n})= n E(\rho)$, are used.

\begin{figure}[t]
\includegraphics[width=0.8\columnwidth]{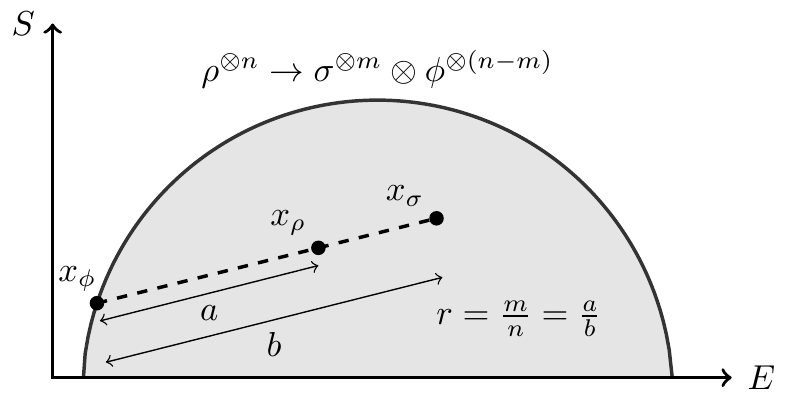}
\caption{
Representation in the energy-entropy diagram and role of operations respecting energy and entropy conservation
constraints in the transformation $\rho^{\otimes n}\to\sigma^{\otimes m}\otimes \phi^{\otimes n-m}$ imply $x_\rho$, $x_\sigma$ and $x_\phi$ to be aligned and $x_\rho$ to lie in between $x_\sigma$ and $x_\phi$.
\label{fig:convex-combination}}
\end{figure}

As shown in Fig.~\ref{fig:convex-combination}, the three points $x_{\rho}, \ x_\sigma$ and $x_{\phi}$ lie on a straight line in the energy-entropy diagram, as a consequence of Eq.~\eqref{eq:convex-combination}. Moreover, $0\le r \le 1$ implies that $x_\rho$ lies in between $x_\sigma$ and $x_\phi$. The conversion rate $r$ can be given a geometric interpretation and it is the Euclidean distance between $x_\phi$ and $x_\rho$ relative to the total Euclidean distance between $x_\sigma$ and $x_\phi$ (see Fig.~\ref{fig:convex-combination}). 

The conversion rate $r$, from $\rho$ and $\sigma$, becomes maximal when the state $\phi$ lies on the boundary of the energy-entropy diagram. That means it would be either a thermal or a pure state. Quantitatively, the conversion rate $r$ is maximized, when 
\begin{equation*}
\frac{S(\sigma)-S(\phi)}{S(\sigma)-S(\rho)}=\frac{E(\sigma)-E(\phi)}{E(\sigma)-E(\rho)},
\end{equation*}
and $\phi$ is thermal state. Accordingly, the rate becomes
\begin{equation}\label{eq:interconvertibility-rate}
r=\frac{m}{n}=\frac{S(\rho)-S(\phi)}{S(\sigma)-S(\phi)}\, .
\end{equation}
This can be easily seen, geometrically, in Fig.~\ref{fig:convex-combination}. The Eq.~\eqref{eq:rate-entropies} is recovered in the case of $\phi$ being pure (vanishing entropy). Let us note that, an alternative derivation of the transformation rate is obtained in \cite{Sparaciari16}. However, they can be shown to be identical. The rate in Eq.~\eqref{eq:interconvertibility-rate} is more compact and less technical, compared to the one in \cite{Sparaciari16}. 

We note that the resource theory of quantum thermodynamics, presented in Chapter \chrto, can be understood as a special case of our consideration above, where system is attached to an arbitrarily large bath at fixed temperature and allowed operations are global energy preserving unitaries, instead of more general global entropy preserving operations.


\section{Discussion}

Thermodynamics, and in particular work extraction from non-equilibrium states, has been studied in the quantum domain, in the recent years. It introduces radically new insights into quantum statistical and thermal processes. In much these studies, be it classical or quantum, thermal baths are assumed to be considerably large in size compared to systems under consideration. That is why, the baths remains always thermal, with same temperature, before and after it interacts with a system. Also, an equilibrated system always shares the same temperature with the bath. Indeed, the assumption large is not fulfilled in every situation. If the baths are finite and small systems, the standard formulation of thermodynamics breaks down. The first problem one would encounter is the inconsistency in the notion of temperature itself. A finite bath could go out of thermal equilibrium, by exchanging energy with a system. Such a situation is relevant for thermodynamics that applies to quantum regime, where system and bath could be small and comparable in size.  
To incorporate such scenarios, we need to develop a temperature independent thermodynamics, where the bath could be small or large and will not have a special status.   

Here, we have introduced temperature independent formulation of thermodynamics as an exclusive consequence of (coarse-grained) information conservation. The information is measured in terms of von Neumann entropy. The formulation is relied on the fact that systems with same entropy can be inter-convertible using entropy preserving operations. Therefore, the states with same entropy forms a constant entropy manifold and there exists a state that possesses minimal amount of energy. This state with minimal energy are known as {\it a completely passive state}, which assumes a  Boltzmann--Gibb's canonical form with an intrinsic temperature. The energy of a completely passive state is defined as the bound energy, as this energy cannot be extracted by any entropy preserving operations. For any given state, the free energy is defined as the difference between the internal energy and the bound energy, as this amount of energy can be accessible by means of entropy preserving operations.  As shown in \cite{Sparaciari16}, two different states possessing identical energy and entropy are thermodynamically equivalent. Such equivalence enables us to exploit energy-entropy diagram to understand bound, free energies geometrically. 


With these machinery, we have introduced a completely new definition of heat in terms of bound energy, applicable for arbitrary systems and without any reference to a temperature. We have formulated the laws of thermodynamics accordingly and, as we have seen, they are a consequence of the reversible dynamics of the underlying physical theory.
In particular:
\begin{itemize}
\item \emph{Zeroth law} is a consequence of information conservation.
\item \emph{First and second laws} are a consequence of energy conservation, together with information conservation.
\item \emph{Third law} is a consequence of "strict" information conservation (i.e. microscopic reversibility or unitarity). There is no third law for processes that only respect "coarse-grained" information conservation. 
\end{itemize} 

We have applied our formalism to the heat engines that consist of finite bath and demonstrated that the maximum efficiency is in general less, compared to  an ideal Carnot's engine. We have also introduced a resource theoretic framework for intrinsic temperature based thermodynamics. This approach enables us to address the problem of inter-state transformations and work extraction. These results are given a geometric meaning, in terms of  the energy-entropy diagram. 

The information conservation based framework for thermodynamics can be extended to multiple conserved quantities \cite{BeraPRX}. Analogously, charge-entropy and resource theory can given in this scenario. The extraction of a generalized potential (i.e. linear combinations of charges),
becomes analogous to the work extraction (the single charge case).

An immediate question arises is that to what extent the formalism can be extended beyond coarse-grained information conservation operations. This is an interesting open question, as in that case, there would be a different notion of bound energy and possibly many more equivalence classes of states.
It is also far from clear if energy-entropy diagrams would be meaningful there. 



\bigskip

\section*{Acknowledgements}

We acknowledge financial support from the European Commission 
(FETPRO QUIC H2020-FETPROACT-2014 No. 641122), 
the European Research Council (AdG OSYRIS and AdG IRQUAT), 
the Spanish MINECO (grants no.  FIS2008-01236, FISICATEAMO FIS2016-79508-P,
FIS2013-40627-P, FIS2016-86681-P,
and Severo Ochoa Excellence Grant SEV-2015-0522) with the support of FEDER funds, 
the Generalitat de Catalunya (grants no.~2017 SGR 1341, and SGR 875 and 966), CERCA  Program/Generalitat  de  Catalunya
and Fundaci{\'o} Privada Cellex.
MNB also thanks support from the ICFO-MPQ fellowship.



\end{document}